\documentclass[12pt]{amsart}
\usepackage{graphicx}
\usepackage{amssymb,amsmath,amsthm,amscd}

\input xy
\xyoption{all}
\UseComputerModernTips

\newcommand{\R}{\mathbb{R}}
\newcommand{\C}{\mathbb{C}}
\newcommand{\Z}{\mathbb{Z}}
\newcommand{\po}{{\mathfrak{p}}}
\newcommand{\poZ}{\mathfrak{p}_{\Z}}
\newcommand{\dt}{\partial_{t}}
\newcommand{\bbS}{\mathbb{S}}
\newcommand{\V}{\mathcal{V}}
\newcommand{\Vk}{V^{(k)}}
\newcommand{\g}{\mathfrak{g}}
\newcommand{\Vsigma}{\V_{\sigma-1}}
\newcommand{\A}{\mathcal{A}}
\newcommand{\M}{\mathcal{M}}
\newcommand{\Asigma}{\A_{\sigma-1}}
\newcommand{\Msigma}{\M_{\sigma-1}}
\newcommand{\semi}{\ltimes}

\DeclareMathOperator{\Gr}{Gr}

\DeclareMathOperator{\Span}{Span}
\DeclareMathOperator{\SO}{SO}
\DeclareMathOperator{\Id}{Id}
\DeclareMathOperator{\Cl}{Cl}
\DeclareMathOperator{\Map}{Map}
\DeclareMathOperator{\Def}{Def}
\DeclareMathOperator{\App}{App}
\DeclareMathOperator{\ev}{ev}
\DeclareMathOperator{\Ker}{Ker}
\DeclareMathOperator{\End}{End}
\newcommand{\Deft}{\Def_{s}}
\newcommand{\Defst}{\Def_{s,t}}
\DeclareMathOperator{\Spin}{Spin}
\DeclareMathOperator{\Sym}{Sym}

\theoremstyle{plain}
\newtheorem{theorem}{Theorem}[section]
\newtheorem{corollary}[theorem]{Corollary}

\newtheorem{lemma}[theorem]{Lemma}

\theoremstyle{remark}
\newtheorem{definition}[theorem]{Definition}
\newtheorem{remark}[theorem]{Remark}
\newtheorem{example}[theorem]{Example}

\numberwithin{equation}{section}

\title{Off-shell supersymmetry and filtered Clifford supermodules}

\subjclass[2000]{Primary: 81Q60; Secondary: 15A66, 16W70}
\keywords{off-shell supersymmetry, supersymmetric quantum mechanics, super Poincar\'e, Clifford algebra, spinor, Adinkra, filtration, bifiltration}
\author[C.~Doran]{Charles F.~Doran}
\address{C.~Doran\\University of Washington\\Department of Mathematics\\Box 354350\\Seattle, WA 98105-4350}
\email{cfd@u.washington.edu}
\urladdr{http://www.math.washington.edu/\~{}doran/}

\author[M.~Faux]{Michael G.~Faux}
\address{M.~Faux\\Department of Physics\\SUNY College at Oneonta, Oneonta, NY 13820}
\email{fauxmg@oneonta.edu}
\urladdr{http://employees.oneonta.edu/fauxmg/}

\author[S.~J.~Gates, Jr.]{S.~James Gates, Jr.}
\address{S.~J.~Gates, Jr.\\University of Maryland\\Physics Department\\Rm.~4125\\College Park, MD 20742-4111}
\email{gatess@wam.umd.edu}
\urladdr{http://www.physics.umd.edu/ep/gates/gates.html}

\author[T.~H\"ubsch]{Tristan H\"ubsch}
\address{T.~H\"ubsch\\Howard University\\Department of Physics \& Astronomy and Department of Mathematics\\2355 Sixth St., NW, Rm. 213\\Washington, DC 20059}
\email{thubsch@howard.edu}
\urladdr{http://string.howard.edu/\~{}tristan/}
\thanks{T.~H\"ubsch is supported by the Department of Energy through the grant DE-FG02-94ER-40854.}
\author[K.~Iga]{Kevin M. Iga}
\address{K.~Iga\\Natural Science Division\\Pepperdine University\\24255 Pacific Coast Hwy.\\Malibu, CA 90263}
\email{Kevin.Iga@pepperdine.edu}
\urladdr{http://math.pepperdine.edu/kiga/}

\author[G.~Landweber]{Gregory D.~Landweber}
\address{G.~Landweber\\Mathematics Department\\University of Oregon\\Eugene, OR 97403-1222}
\email{greg@uoregon.edu}
\urladdr{http://math.uoregon.edu/\~{}greg/}

\date{\today}
\begin{document}

\begin{abstract}
An off-shell representation of supersymmetry is a representation of the
super Poincar\'e algebra on a dynamically unconstrained space of fields. We describe
such representations formally, in terms of the fields and their spacetime derivatives, and we interpret the physical concept of engineering dimension
as an integral grading. We prove that formal graded off-shell representations of one-dimensional $N$-extended supersymmetry, i.e., the super Poincar\'e algebra $\po^{1|N}$, correspond to filtered Clifford supermodules
over $\Cl(N)$. We also prove that formal graded off-shell representations of two-dimensional $(p,q)$-supersymmetry, i.e., the super Poincar\'e algebra $\po^{1,1|p,q}$, correspond to bifiltered Clifford supermodules over $\Cl(p+q)$.

Our primary tools are the formal deformations of filtered superalgebras and supermodules, which give a one-to-one correspondence between filtered spaces
and graded spaces with even degree-shifting injections. This generalizes the machinery developed by Gerstenhaber to prove that every filtered algebra is a deformation of its associated graded algebra. Our treatment extends Gerstenhaber's discussion to the case of filtrations which are compatible with a supersymmetric structure, as well as to filtered modules in addition to filtered algebras. We also describe the analogous constructions for bifiltrations and bigradings.
\end{abstract}
\maketitle

\section{Introduction}

In mathematics, the term ``supersymmetry'' is used to describe algebraic structures which possess a $\Z_{2}$-grading and obey standard sign conventions related to that grading. In physics, the term ``supersymmetry'' is much more specific, referring to structures which are equivariant with respect to  the super Poincar\'e group, the super Poincar\'e algebra, or their many variants. The Poincar\'e group is the Lie group of isometries of Minkowski space, or more precisely its double cover, replacing $\SO(1,d-1)$ with $\Spin(1,d-1)$.
The super Poincar\'e group is the Lie supergroup obtained by extending the Poincar\'e group by infinitesimal odd elements, called supersymmetry generators, which square to spacetime derivatives, the infinitesimal generators of translations. At the Lie algebra level, the supersymmetry generators span the odd component of the super Poincar\'e algebra.

The physical representations of the super Poincar\'e group and super Poincar\'e algebra come in two forms. Both are representations on spaces of fields, i.e., maps from Minkowski space to a finite-dimensional $\Z_{2}$-graded representation of $\Spin(1,d-1)$. The $\Z_{2}$-grading decomposes the fields into bosons and fermions, and the Lorentz action decomposes the fields into irreducible components, each corresponding to a different type of particle.
The assembly of several such particles into a representation of supersymmetry is called a supermultiplet. The Poincar\'e group acts naturally on such spaces of fields, and the question which remains is how the supersymmetry generators in the super Poincar\'e algebra will act.

In off-shell representations, the super Poincar\'e algebra acts on dynamically unconstrained spaces of fields, while on-shell representations restrict the action to fields which satisfy the equations of motion, usually coming from a  Lagrangian via the Euler-Lagrange equations. Although on-shell representation
are more complicated physically, they are more natural from the point of view of representation theory. In particular, the irreducible unitary representation of the super Poincar\'e algebra are all on-shell representations, which can be classified via a supersymmetric version of Wigner's methods (see, for example, \cite{Freed99}). In off-shell representations, the supersymmetry is manifest from the description of the particles in the supermultiplet, allowing us to separate the representation theory from the physics, i.e., the Lagrangian, and facilitating quantization.
However, most supersymmetric theories, including covariant fields for string theory and $M$-theory, are known only in on-shell formulations, in which the action of the supersymmetry generators satisfies the relations of the super Poincar\'e algebra only once we additionally impose the equations of motion.

Given an off-shell representation, one can restrict it to obtain various on-shell representations corresponding to different masses and Lagrangians. In
\cite{Gates02}, Gates \textit{et al.} remind the high energy physics community that the ``fundamental supersymmetry challenge'' is to reverse this process, showing when and how a given on-shell representation admits an off-shell extension. In order to address this challenge, we are working towards a classification of off-shell supersymmetric theories, a problem which has been left unsolved for over 30 years. In \cite{Gates02a}, Gates \textit{et al.} propose studying off-shell supersymmetric representations in terms of their dimensional reductions to one time-like dimension, noting that off-shell representations of one-dimensional supersymmetry can be constructed using Clifford algebras. The purpose of this article is to give a mathematically rigorous study of the role of Clifford algebras in low-dimensional off-shell supersymmetry.

The main result of this article is that off-shell supersymmetric representations for one-dimensional and two-dimensional Minkowski space correspond to filtered and bifiltered Clifford supermodules, respectively. More precisely, off-shell representations of the $d=1$, $N$-extended super Poincar\'e algebra correspond to filtered $\Cl(N)$-supermodules, while off-shell representations of the $d=2$, $(p,q)$-extended super Poincar\'e algebra correspond to bifiltered $\Cl(p+q)$-supermodules. Under this correspondence, the filtrations on these Clifford supermodules give rise to gradings on the off-shell representations, which encode the physical concept of engineering dimension. In order to extend a representation of the Clifford algebra to an off-shell representation of the one-dimensional super Poincar\'e algebra, one needs to additionally specify the action of the infinitesimal generator $\partial_{t}$ of time translations. Our arguments show that this additional information corresponds precisely to a choice of filtration on the Clifford supermodule. In two dimensions, Minkowski space decomposes into a sum of two complementary Lorentz invariant light-cone subspaces. To extend a representation of the Clifford algebra to an off-shell representation of the two-dimensional super Poincar\'e algebra, we therefore require two compatible filtrations, which allow us to specify the action of the infinitesimal generators of translations along each of these two light-cone subspaces.

In Section 2, we describe off-shell and on-shell representations algebraically rather than analytically, viewing the fields and their derivatives as formal objects, rather than as functions on Minkowski space. We introduce supersymmetric filtrations in Section 3, and in Section 4 we generalize the machinery which Gerstenhaber developed in \cite{Ger66} to prove that every filtered algebra is a deformation of its associated graded algebra. We describe formal deformations of filtered spaces, extending Gerstenhaber's discussion to filtrations which are compatible with the supersymmetric structure, and to formal deformations of filtered modules as well as filtered algebras. We use these results in Section 5, where we prove that the universal enveloping superalgebra of the one-dimensional super Poincar\'e algebra is the formal deformation of the Clifford algebra with respect to its standard filtration. It then follows that formal graded off-shell representations of one-dimensional supersymmetry are the formal deformations of filtered Clifford supermodules. In Section 6, we reprise the discussion of Sections 3 and 4
for bifiltrations and bigradings, which allows us to prove in Section 7 that
formal graded off-shell representation of two-dimensional supersymmetry are the formal deformations of bifiltered Clifford supermodules.

The authors would like to thank Aravind Asok, who pointed out the reference \cite{Ger66}.

\section{Off-shell representations}

\subsection{The super Poincar\'e algebra}

In non-supersymmetric field theory one considers theories which are invariant under the action of the Poincar\'e group
$$P^{d} = \Spin(1,d-1) \semi \R^{1,d-1}$$
of isometries of Minkowski space. Here $\Spin(1,d-1)$ acts by Lorentz transformations and $\R^{1,d-1}$ acts by translations of Minkowski space. Infinitesimally, the translations are generated by spacetime derivative operators.  In supersymmetric field theory, one extends the Lie algebra of the Poincar\'e group by introducing odd operators, the supersymmetry generators $Q_{i}$, whose squares are spacetime derivatives.

We first consider the special case where $d=1$, with arbitrarily many odd supersymmetry generators all squaring to the single time-like derivative operator $\partial_{t}$, denoted formally by $H$. This setting corresponds to the infinitesimal symmetries of supersymmetric quantum mechanics. Since the group $\Spin(1) \cong \Z_{2}$ is discrete, there are no infinitesimal generators of Lorentz symmetry here.

\begin{definition}\label{def:poincare}
	The $d=1$, $N$-extended \textit{super Poincar\'e algebra} is the real
	Lie superalgebra
	$$\po^{1|N} = \R H \oplus \Pi \Span\{Q_{1},\ldots,Q_{N}\},$$
	with even generator $H$ and odd
	generators $Q_{1},\ldots,Q_{N}$ satisfying
	$$\{Q_{i},Q_{j}\} = 2\,\delta_{ij} H$$
	for $i,j = 1,\ldots,N$.
\end{definition}
Here we use the parity reversal operator $\Pi$ to remind ourselves that $\Pi\Span\{Q_{1},\ldots,Q_{N}\}$ is an odd vector space. We note that it follows from the definition that $[H,H] = 0$ and $[H,Q_{i}] = 0$ for all $i = 1,\ldots,N$. More generally, we can replace the span of the odd generators with an inner product space $\bbS$, defining
\begin{equation}\label{eq:bbS1}
	\po(1,\bbS) = \R H \oplus \Pi \bbS
\end{equation}
with brackets
\begin{equation}\label{eq:bbS2}
	\{s,t\} = 2\langle s,t\rangle H
\end{equation}
for $s,t\in \Pi\bbS$, where $\langle\cdot,\cdot\rangle$ is a positive definite symmetric bilinear form on $\bbS$.

\subsection{Engineering dimension}

In physical systems, measurements usually come with units.  For instance, a mass is not simply a number, but rather a quantity such as $9.11\times 10^{-31}$ kilograms. The International System of Units (SI) consists of basic units, such as meter, kilogram, second, kelvin, ampere, and so on, and derived units, such as the Joule ($= \mbox{kilogram}\cdot \mbox{meter}^2/\mbox{second}^2$).
In fact, we can use fundamental constants to eliminate most of the basic units.  For instance,
multiplying measurements of temperature
in kelvin by Boltzmann's constant $k \cong 1.380\times 10^{-23}$ Joule/kelvin we can express temperatures not in kelvins, but rather in terms of kilograms, meters, and seconds.
Equivalently, we can replace kelvin with a unit of temperature such that Boltzmann's constant is 1.
Similarly, we can define a unit of charge so that the constant $k$ in Coulomb's law $F=\frac{k q_1 q_2}{r^2}$ is 1, and thus eliminate amperes.  We can eliminate meters by setting the speed of light $c=1$.  Planck's constant $\hbar$ in quantum mechanics then has units $\mbox{kilogram}\cdot \mbox{second}$, and by setting $\hbar = 1$ we can eliminate seconds.
The only basic unit remaining is that of mass, expressed as kilograms in SI units, or more commonly as electron-volts$/c^{2}$ in the particle physics literature.

Writing $[M]$ to indicate our choice of units for mass, all
other units will then be of the form $[M]^n$ for some $n$.  The exponent $n$ is called the \textit{engineering dimension} (also referred to as mass dimension or canonical dimension).
In these units, time is measured in units of $[M]^{-1}$, and thus the derivative operator $d/dt$ carries units of $[M]$.  Since the supersymmetry generators $Q_i$ correspond to square roots of derivatives, the units of $Q_i$ should be $[M]^{1/2}$. In other words, the generator $H$ has engineering dimension 1, while the supersymmetry generators $Q_{i}$ have engineering dimension $1/2$.

Fields can also be expressed in units of $[M]^n$.
The action is dimensionless, with units $[M]^0$, and thus a Lagrangian in $d$ dimensions has units $[M]^d$.  Bosonic fields typically enter the Lagrangian via terms of the form $k\,\|d \phi\|^2$ or $k\,\|\phi\|^2$, where $k$ has integral engineering dimension, while fermionic fields typically enter the Lagrangian via terms of the form $k\,\langle \psi, d\psi\rangle$.  When $d$ is even, bosonic fields have integral engineering dimension and fermionic fields have half-integral engineering dimension, and when $d$ is odd, the reverse is true. 
From a mathematical perspective, the engineering dimension corresponds to an integral grading on the super Poincar\'e algebra and spaces of fields.  For notational convenience, and to align this terminology with mathematical conventions for integral gradings, we define the degree to be twice the engineering dimension, replacing ``integral'' and ``half-integral'' engineering dimensions with ``even'' and ``odd'' degrees, respectively.

\begin{definition}
	The \textit{graded super Poincar\'e algebra} is the $\Z$-graded real Lie superalgebra
	$$\poZ^{1|N} = \R H_{(2)} \oplus \Pi \Span\{ Q_{1},\ldots,Q_{N}\}_{(1)},$$
	with the same underlying Lie superalgebra as $\po^{1|N}$, but with the
	generators graded by:
	$$\deg H = 2, \quad
	\deg Q_{1} = \cdots = \deg Q_{N} = 1.$$
\end{definition}

\subsection{Off-shell and on-shell fields}
Physicists consider two distinct types of representations of the Poincar\'e algebra and the super Poincar\'e algebra. Both act on fields, spaces of functions (or sections) the form
\begin{equation}\label{eq:H-field}
	\mathcal{H} = \Map\bigl(\R^{1,d-1},V\bigr),
\end{equation}
where $V$ is a representation of $\Spin(1,d-1)$. If we are considering representations of the super Poincar\'e algebra or graded super Poincar\'e algebra, then $V$ is furthermore $\Z_{2}$-graded or $\Z$-graded, respectively, in order to keep track of bosons and fermions and their engineering dimensions. The space $\mathcal{H}$ carries a natural action of the Poincar\'e group and the Poincar\'e algebra, with $\Spin(1,d-1)$ acting by conjugation and $\R^{1,d-1}$ acting by translation. The problem of finding representations of supersymmetry therefore amounts to determining how the supersymmetry generators $Q_{i}$ can act so as to square to the generators of translations.

In an \textit{off-shell} representation, we have no dynamical conditions on our fields.  On the other hand, in an \textit{on-shell} representation, we restrict our attention to the subset of fields $\phi(x)$ satisfying the equations of motion, usually related to the Klein-Gordon equation
$$ (\Box  - m^{2})\,\phi(x) = 0.$$
Here $\Box$ is the d'Alembertian operator, i.e., the Minkowski space Laplacian, and $m$ is the mass, a non-negative real constant.
If $\hat{\phi}$ is the Fourier transform of our field $\phi$, then the Klein-Gordon equation becomes
$$ (|k|^{2} - m^{2})\,\hat{\phi}(k) = 0, $$
and the solution set consists of all fields $\phi(x)$ whose Fourier transforms $\hat{\phi}(k)$ are supported along the \textit{mass shells}
$$\mathcal{O}_{m} = \{ k \in (\R^{1,d-1})^{*} : |k|^{2} = m^{2} \}$$
These shells are generally hyperboloids, with the exception of the trivial shell at 0 and the light-cone, and correspond to the covector orbits of $\Spin(1,n-1)$ acting on momentum space $(\R^{1,d-1})^{*}$. 
Given an off-shell representation $\mathcal{H}$ as in (\ref{eq:H-field}),
we obtain on-shell representations for each mass shell $\mathcal{O}_{m}$,
$$\mathcal{H}_{m} = \bigl\{ \phi(x) \in \mathcal{H} : \hat{\phi}(k) = 0 \text{ for } |k|^{2}\neq m^{2} \bigr\},$$
by restricting $\mathcal{H}$ to those fields whose Fourier transforms are supported on $\mathcal{O}_{m}$.

When $d = 1$, a shell is simply a single point $k$, and the corresponding
on-shell fields are functions of the form $f(t) = e^{i kt}\,v$ with $v\in V$. The even generator $H = \partial_{t}$ of the super Poincar\'e algebra then acts on this on-shell representation by the constant $ik\,\Id$, as
\begin{equation}\label{eq:on-shell-physics}
H f(t) = \partial_{t}\, e^{ikt}\,v = ik\,e^{ikt}\,v = ik\,f(t)
\end{equation}
for fields on the shell $k$. (Note that the standard physics convention is to take $H = i \partial_{t}$, in which case the eigenvalues of $H$ are real-valued and correspond to energy).
In this case, a $d=1$ off-shell representation can be viewed as a smooth family of $d=1$ on-shell representations parameterized by the shells $k \in \R$.

\subsection{Formal representations of supersymmetry}
We now give formal algebraic definitions of the physical concepts of on-shell and off-shell representations. We begin with on-shell representations, as they are simpler to describe.

\begin{definition}
A \textit{formal on-shell representation} of the super Poincar\'e algebra
is a $\po^{1|N}$-supermodule on which the even generator $H$ acts by a multiple of the identity operator.
\end{definition}

From (\ref{eq:on-shell-physics}), we see that in a physical on-shell representation, the generator $H$ acts by an imaginary multiple of the identity. By our definition, we can also consider formal on-shell representations in which $H$ acts by a real multiple of the identity,
and in particular our definition of on-shell does not require any complex structures. Since the generator $H$ is central, it follows from Schur's lemma that $H$ must act by a constant multiple of the identity on any complex irreducible $\po^{1|N}$-supermodule. Therefore, all irreducible complex $\po^{1|N}$-supermodules are on-shell representations.

On an off-shell representation $\mathcal{H}$ given by (\ref{eq:H-field}), the even generator $H$ acts by the operator $\dt$ generating time translations. Physicists typically require the odd generators $Q_{i}$ to act by local operators, which by Peetre's theorem implies
that they act by differential operators. We can therefore write
the action of each supersymmetry generator in the form
$$Q \mapsto \sum_{n \geq 0} a_{n}\,{\dt}^{n},$$
and since we have $[H,Q] = 0$, the coefficients $a_{n} \in \End V$
are constant in $t$. The supersymmetry generators $Q_{i}$ thus act on an off-shell representation by elements of the superalgebra $\End V \otimes \R[\dt]$. Returning to the even generator $H$ acting by $\dt$, we note that the algebra $\R[H]$ acts faithfully on unconstrained off-shell field by elements of $\Id_{V} \otimes \R[\dt]$. 
This leads us to the following definition:

\begin{definition}\label{def:off-shell}
	A \textit{formal off-shell representation} of the graded super Poincar\'e algebra is a $\Z$-graded representation of $\poZ^{1|N}$
on a finitely generated, free $\R[H]$-supermodule.
\end{definition}

In other words, a formal off-shell representation is of the form
$$\V \cong V \otimes \R[H],$$
where $V = \bigoplus_{j\in \Z} V_{j}$ is a finite-dimensional $\Z$-graded vector space.  
Note that we could alternatively define a formal off-shell representation as a $\Z$-graded representation of $\poZ^{1|N}$ on which the even generator $H$ acts injectively.
Given a formal off-shell representation $\V$, we can extract the corresponding finite-dimensional $\Z$-graded vector space $V$ by taking the quotient
\begin{equation}\label{eq:V}
V := \V \,/\, H \V.
\end{equation}
Since $V$ is finite-dimensional and $\deg H = +2$, we note that such a representation necessarily has a minimum overall degree, given by the minimum degree appearing in $V$, which we typically normalize to be $0$ or $1$, depending whether the lowest degree subspace has even or odd degree.

Using this definition of a formal off-shell representation, we view our fields not as functions, but rather as abstract symbols valued in the finite-dimensional vector space $V$, graded according to the engineering dimension. This viewpoint is implicit in much of the physics literature. Indeed, when physicists talk about
a ``basis'' for fields, they typically do not mean an infinite-dimensional basis for a space of functions, but rather a basis for the finite-dimensional vector space $V$. From this point of view,
the derivatives of fields give additional degrees of freedom within an off-shell representation. For instance, a Lagrangian is a formal expression built from the fields and their derivatives, treated as independent parameters.

\begin{remark}
We can, in fact, embed any formal off-shell representation as a dense subspace
of a Hilbert space of fields $\mathcal{H} = L^{2}(\R,V)$. Let $\Phi(t) = e^{-t^{2}/2}$ denote the Gaussian function. Given a formal off-shell representation $\V \cong V \otimes \R[H]$, we can construct an embedding 
$$v \otimes p(H) \in \V \longmapsto p(\partial_{t})\,\Phi(t) v \in \mathcal{H},$$
for a vector $v\in V$ and a polynomial $p(H)\in \R[H]$. Here we treat the Gaussian as a ``vacuum'' state in $L^{2}(\R)$ and build a Fock space
isomorphic to $\V$ by acting on $\Phi \otimes V$ by the operator $\partial_{t} \otimes 1$.
This is clearly a homomorphism of $\R[H]$-modules, where $H$ acts by $\partial_{t}\otimes 1$ on $\mathcal{H}$. The kernel of this map is trivial since the algebra $\R[\partial_{t}]$ acts freely on the ``vacuum'' $\Phi(t)$. Furthermore, recalling that
$\R[\partial_{t}]\,\Phi(t) = \R[t]\,\Phi(t)$, we observe that the image of $\V$ is dense in $\mathcal{H}$, since any smooth function, and consequently any element of the Hilbert space $L^{2}(\R)$, can be approximated by a polynomial.
\end{remark}

Given a formal off-shell representation, we can construct corresponding formal on-shell representations for any shell $k\in \R$. Perturbing equation (\ref{eq:V}) for the finite-dimensional $\Z$-graded vector space $V$, we define
\begin{equation}\label{eq:Vk}
\V_{H-k} := \V \otimes_{\R[H]} \R_{k}
\cong \V \,/\, (H - k \Id)\V,
\end{equation}
where $\R_{k}$ is the one-dimensional representation of $\R[H]$ on which the generator $H$ acts by multiplication by the constant $k\in \R$. Since $\deg H = 2$ while $\deg \Id = 0$, the ideal generated by $H-k\cdot \Id$ is inhomogeneous (one could make it homogeneous by inserting the appropriate units for $k$ or the appropriate power of $\hbar$), and thus the quotient $\Vk$ no longer possesses a $\Z$-grading, that is unless $k = 0$, in which case $\V_{H=0} = V$. On the other hand, since both $H$ and $\Id$ have even degree, we obtain a residual $\Z_{2}$-grading, making $\V_{H-k}$ a super vector space. In addition, we obtain a residual integral filtration on $\V_{H=k}$, which we discuss in the following section.

\section{Filtered supermodules}\label{sec:filtrations}

We begin this section with a few algebraic definitions.

\begin{definition}\label{def:filtration}
	An (increasing) filtration on a super vector space $V = V_{0} \oplus V_{1}$ is a 
	collection of even subspaces $F_{2q}(V) \subset V_{0}$ for $q\in\Z$ and odd subspaces $F_{2q+1}(V) \subset V_{1}$ for $q\in \Z$ such that $F_{p}(V) \subset F_{p+2}(V)$ for all $p\in\Z$.
	We normalize filtrations so that $F_{p}(V) = 0$ for $p < 0$.
\end{definition}

In other words, a filtration of a super vector space consists of two separate filtrations for the even and odd degree components. Alternatively, we could consider
a more conventional (increasing) filtration of $\Z_{2}$-graded subspaces $F'_{p}(V) \subset V$ for $p \in \Z$ with $F'_{p}(V) \subset F'_{p+1}(V)$ for all $p\in \Z$, together with the supersymmetric condition
$$F'_{p}(A)_{p \bmod 2} = F'_{p+1}(A)_{p \bmod 2} \text{ for all } p\in \Z.$$
These two types of super filtrations are related to each other by taking
$$F'_{p}(V) = F_{p}(V) \oplus F_{p-1}(V) \text{ and } F_{p}(V) = F'_{p}(V)_{p\bmod 2}.$$ In the following, we use the convention given in Definition~\ref{def:filtration}.

\begin{definition}
	A \textit{filtered superalgebra} is a $\Z_{2}$-graded algebra $A = A_{0}\oplus A_{1}$ together with a  filtration on $A$ satisfying the multiplicative property
	$$F_{p}(A) \cdot F_{q}(A) \subset F_{p+q}(A)$$
for all $p,q\in \Z$.
\end{definition}

\begin{example}
	Let $\g = \g_{0} \oplus \g_{1}$ be a Lie superalgebra. Then the universal enveloping superalgebra $U(\g)$ is a filtered superalgebra. Since $U(\g)$ is generated as an algebra by the elements of the Lie superalgebra $\g$, we obtain a super filtration on $U(\g)$ by assigning the odd component $\g_{1}$ filtration degree 1 and the even component $\g_{0}$ filtration degree 2. We observe that the relations for the universal enveloping superalgebra,
	$$X Y - (-1)^{|X|\,|Y|} Y X = [X,Y]$$
for $X,Y\in \g$ homogeneous elements of $\Z_{2}$-degrees $|X|$ and $|Y|$, respectively, respect this super filtration. 
\end{example}

\begin{definition}
	A \textit{filtered supermodule} over a filtered superalgebra $A$ is a supermodule $V$ over $A$ together with a filtration satisfying the multiplicative property
	$$F_{p}(A) \cdot F_{q}(V) \subset F_{p+q}(V)$$
for all $p,q\in \Z$.
\end{definition}

In particular, by comparing these two definitions, we see that a filtered superalgebra is a filtered supermodule over itself.

\begin{definition}
Let $\V$ be a $\Z$-graded vector space, algebra, or module. The $\Z$-grading on $\V$ then induces a natural increasing super filtration on $\V$ by degree, given by
\begin{equation}\label{eq:natural-filtration}
F_{p}\V := \bigoplus_{\substack{j\leq p\\j\equiv p\bmod 2}} \V_{j},
\end{equation}
with respect to the $\Z_{2}$-grading 
$\V = \V_{\text{even}} \oplus \V_{\text{odd}}.$ If $\V_{p} = 0$ for $p < 0$, then $F_{p}\V = 0$ for $p < 0$. If $\A$ is a graded algebra, this construction gives $\A$ the structure of a filtered superalgebra. A filtered supermodule over a graded algebra $\A$ is then a filtered supermodule with respect to the natural induced super filtration on $\A$.
\end{definition}

We now return to representations of the super Poincar\'e algebra.
Given a formal graded off-shell representation $\V$, we can view it as a filtered $\poZ^{1|N}$-supermodule provided that we filter $\V$ according to its integral degree by (\ref{eq:natural-filtration}). For any fixed shell $k\in \R$, we can project the filtration on the formal off-shell representation $\V$ down to the corresponding formal on-shell representation $\V_{H-k}$ given by (\ref{eq:Vk}) by taking
$$F_{p}\V_{H-k} = F_{p}\V \,/\, (H-k\Id)F_{p-2}\V.$$
This gives us a finite increasing filtration on $\V_{H-k}$, giving the formal on-shell representation $\V_{H-k}$ the structure of a filtered supermodule over $\po_{Z}^{1|N}$.

\section{Deformations of filtered supermodules}\label{sec:deformations}

The discussion in this section is based on \cite{Ger66}. Our treatment differs in that we consider increasing rather than decreasing filtrations, we consider superalgebras rather than algebras, and we extend the discussion to supermodules. Although we work here with vector spaces and algebras over a field (which we will later take to be $\R$ or $\C$), one can just as easily modify our discussion to consider modules and algebras over an arbitrary coefficient ring $k$.

\begin{definition}
	Given a (increasingly) filtered super vector space $V = V_{0} \oplus V_{1}$,
	its formal deformation $\Deft V$ is the subspace of $V[s]$ consisting of
	finite polynomials of the form
	$$\Deft V := \left\{ {\sum}_{p\in\Z} v_{p}\, s^{p} \in V[s] : v_{p}\in F_{p}V \right\}.$$
	The formal deformation $\Deft V$ is naturally $\Z$-graded by powers of $s$.
\end{definition}

In other words, given a filtered space $V$, we construct a corresponding graded space $\Deft V$ (not to be confused with the associated graded space) whose homogeneous graded subspaces are the filtration levels of $V$:
\begin{equation}\label{eq:grading-is-filtration}
(\Deft V)_{p} \cong F_{p}V.
\end{equation}
We introduce the formal parameter $s$ in order to keep track of the grading on $\Deft V$.

Going in the other direction, let $\V$ be a $\Z$-graded vector space. This grading induces a natural (increasing) super filtration on $\V$ given by (\ref{eq:natural-filtration}).
A \textit{shift} $\sigma \in \End(\V)$ is an (even) endomorphism such that $\sigma \V_{p}\subset \V_{p+2}$. In terms of the filtration, this implies that $\sigma F_{p}\V \subset F_{p+2}\V$.  Let $(\sigma - 1)\V$ denote the subspace of $\V$ given by the image of the map $\sigma - 1 : x \mapsto \sigma x - x$ for all $x \in \V$.  The quotient 
$$\Vsigma := \V / (\sigma - 1)\V$$
is no longer $\Z$-graded, but since the operator $\sigma-1$ preserves the degree mod $2$, we obtain a $\Z_{2}$-grading, making $\Vsigma$ a super vector space.
We also have the compatible (increasing) filtration given by
\begin{equation}\label{eq:filtration-Vprime}
F_{p}(\Vsigma) := F_{p}\V\,/\, (\sigma-1)F_{p-2}\V,
\end{equation}
the image of the natural induced filtration (\ref{eq:natural-filtration}) under the projection $\V \twoheadrightarrow \Vsigma.$

If in addition the shift $\sigma$ is injective, then we can use it to identify $\V_{p}$ with its image $\sigma \V_{p}\subset \V_{p+2}$. Taking the quotient by the subspace $(\sigma - 1)\V$, we obtain by a telescoping construction (see the discussion leading to (\ref{eq:two-filtrations}) below) that
\begin{equation}\label{eq:filtration-is-grading}
F_{p} \Vsigma  \cong \V_{p}.
\end{equation}
This quotient construction therefore takes a graded space with an injective shift and turns its homogeneous graded components into the filtration levels of a corresponding filtered space.

Taking $\V = \Deft V$, we can consider the shift $\sigma = s^{2}$.  We observe
that multiplication by $s^{2}$ on $\Deft V$ is injective. Considering the corresponding filtered super vector space,
\begin{equation}\label{eq:Vprime}
\Def_{1}V := (\Deft V)_{s^{2}=1} = (\Deft V)\,/\,(s^{2}-1)(\Deft V),
\end{equation}
we show in the following lemma that $\Def_{1}V \cong V$, and that in general
these formal deformation and quotient constructions are inverses of each other.

\begin{lemma}
If $V$ is a filtered super vector space, then the filtered super vector space $\Def_{1}V$ given by (\ref{eq:Vprime}) is canonically isomorphic to $V$.  Conversely, if $\V$ is a graded vector space with an injective degree 2 shift
$\sigma$, then $\Deft (\Vsigma)$ is isomorphic to $\V$, identifying $\sigma$ with $s^{2}$.
\end{lemma}
\begin{proof}
It follows from (\ref{eq:grading-is-filtration}) and (\ref{eq:filtration-is-grading})
that $\Def_{1}V \cong V$ and $\Deft(\Vsigma)\cong \V$. Here we construct these canonical isomorphism explicitly, and we describe in detail the ``telescoping construction'' which gives (\ref{eq:filtration-is-grading}). We will require this machinery for the proofs of Lemma~\ref{lemma:def1}
and Lemma~\ref{lemma:shift} below.

Let $V$ be a filtered super vector space and consider the evaluation homomorphism 
\begin{equation}\label{eq:ev1}
\ev_{1} : \sum_{p} v_{p}\,s^{p} \in \Deft V \longmapsto \sum_{p} v_{p} \in V,
\end{equation}
where $v_{p}\in F_{p}V$. This map is clearly surjective, and
we will show that its kernel is the subspace $(s^{2}-1)\Deft V$. Since $V_{p}\subset V_{p+2}$ for all $p$, we observe
that if
$$\ev_{1} : x = v_{0} s^{0} + v_{1} s^{1} + \cdots + v_{n-1} s^{n-1} + v_{n} s^{n} \longmapsto 0,$$
then $v_{n} \in F_{n-2}V$ and $v_{n-1} \in F_{n-3}V$, and thus we can rewrite $x$ in the form
$$x = x' + ( s^{2}-1 ) ( v_{n-1} s^{n-3} + v_{n} s^{n-2} ),$$
where $x'$ is the lower degree polynomial
$$ x' = v_{0} s^{0} + v_{1} s^{1} + \cdots + ( v_{n-3} + v_{n-1} ) s^{n-3}
+ ( v_{n-2} + v_{n}) s^{n-2}.$$
Repeating this process for $x'$, it follows by induction that
\begin{equation}\label{eq:x-in-kernel}
x = ( s^{2} - 1 ) \Biggl( \sum_{p}  \sum_{\substack{j \leq p\\j\equiv p\bmod 2}} v_{j} \,s^{p-2} \Biggr),
\end{equation}
and thus $x\in (s^{2}-1)\Deft V$. Conversely, if $x\in (s^{2}-1)\Deft V$,
then clearly $\ev_{1}(x) = 0$. We therefore obtain $\Ker \ev_{1} = (1-s^{2})\Deft V$,
and thus $\ev_{1}$ descends to a canonical isomorphism $\Def_{1}V = \Deft V / (1-s^{2})\Deft V \cong V$. 

Restricting the evaluation map $\ev_{1}$ to either purely even or purely odd polynomials of degree at most $n$,
we obtain a homomorphism 
$$\ev_{1} : F_{n}(\Deft V) \mapsto F_{n}V,$$
and it follows from (\ref{eq:x-in-kernel}) that the kernel is $(s^{2}-1)F_{n-2}(\Deft V)$. The canonical isomorphism therefore identifies the filtration (\ref{eq:filtration-Vprime}) on $\Def_{1}V$ with the original filtration on $V$.

Conversely, let $\V$ be a graded vector space with an injective degree 2 shift $\sigma$.  Let $\pi : \V \to \Vsigma$ be the projection onto the quotient by $(\sigma-1) \V$.  Given $v_{p} \in \V_{p}$, we have $\pi(v_{p}) \in F_{p}\Vsigma$.  We may therefore consider the map
\begin{equation}\label{eq:f}
f : \sum_{p}v_{p}\in \V \longmapsto \sum_{p}\pi( v_{p}) \,s^{p} \in \Deft(\Vsigma).
\end{equation}
This map is clearly a homomorphism of graded vector spaces.
Suppose that $f ( \sum_{p}v_{p}) = 0$. It follows that $\pi(v_{p}) = 0$ for all $p$. However, we note that $\V_{p} \cap (\sigma - 1)\V  = 0$ for all $p$, and thus all $v_{p}$ must vanish. The map $f$ is therefore injective. To establish
the surjectivity of $f$, we must show that the maps
$\pi_{p} : \V_{p} \to F_{p}\Vsigma$ are surjective for all $p$.
Given any $w \in F_{p}\Vsigma$, it is $\pi(v)$ for some $v\in F_{p}\V$, which by (\ref{eq:natural-filtration}) can be written in the form
$$v = \sum_{\substack{j \leq p\\j \equiv p \bmod 2}} v_{j}$$
However, applying the identity $\pi \circ \sigma = \pi$, we find that
$w = \pi(v) = \pi(v')$, where
$$v' = \sum_{\substack{j \leq p\\j \equiv p \bmod 2}} \sigma^{(p-j)/2}v_{j} \in \V_{p}.$$
In other words, we have
\begin{equation}\label{eq:two-filtrations}
F_{p}\Vsigma = \pi (F_{p}\V) = \pi (\V_{p}).
\end{equation}
It follows that the maps $\pi_{p}$, and in turn the map $f$, are surjective.
\end{proof}

We can also consider formal deformations of superalgebras and supermodules.

\begin{lemma}\label{lemma:def1}
	Let $A$ be a filtered superalgebra, and let $M$ be a filtered supermodule over $A$.
	\begin{enumerate}
		\item
		The formal deformation $\Deft A$ is a $\Z$-graded subalgebra of $A[s]$.
		\item
		The formal deformation $\Deft M \subset M[s]$ is a $\Z$-graded module
		over $\Deft A$.
		\item
		Via (\ref{eq:Vprime}), $\Def_{1}A$ is a filtered superalgebra
		canonically isomorphic to $A$.
		\item
		Via (\ref{eq:Vprime}), $\Def_{1}M$ is a filtered $\Def_{1}A$-supermodule
		canonically isomorphic to $M$.
	\end{enumerate}
\end{lemma}
\begin{proof}
	Since $A$ is a filtered supermodule, we have
	$$(\Deft A)_{p} \, (\Deft A)_{q}
	= F_{p}A \,s^{p} \,F_{q}A \,s^{q}
	\subset F_{p+q}A \,s^{p+q}
	= (\Deft A)_{p+q}.$$
	Likewise, if $M$ is a filtered supermodule over $A$, then
	$$(\Deft A)_{p} \,(\Deft M)_{q}
	= F_{p}A \,s^{p} \,F_{q}M \,s^{q}
	\subset F_{p+q}M \,s^{p+q}
	= (\Deft M)_{p+q},$$
	and thus $\Deft M$ is a graded supermodule over $\Deft A$.
	
	In order to show that $\Def_{1}A$ is canonically isomorphic to $A$, we first note
	that $(s^{2} - 1)\Deft A$ is an ideal in $\Deft A$, thereby giving the
	quotient $\Def_{1}A$ the structure of a filtered superalgebra.
	Secondly, we note that the canonical evaluation homomorphism
	$\ev_{1} : \Deft A \to A$ given by (\ref{eq:ev1})
    is a homomorphism of filtered superalgebras,
	which therefore descends to a filtered superalgebra isomorphism
	$\Def_{1}A \cong A$.
	
	Similarly, in order to show that $\Def_{1}M$ is canonically isomorphic to $M$,
	we first note that $(s^{2} - 1)\Deft M$ is an invariant $\Deft A$-submodule
	of $\Deft M$, which gives the quotient $\Def_{1}A$ the structure of
	a filtered $\Deft A$-supermodule. Secondly, we note that the canonical
	evaluation isomorphism $\ev_{1} : \Deft M \to M$ is a homomorphism of
	$\Deft A$-modules, where $\Deft A$ acts on the filtered $A$-supermodule $M$
	via the homomorphism $\ev_{1} : \Deft A \to A$. We therefore obtain
	an isomorphism $\Def_{1}M \cong M$ as filtered $A$-supermodules.
\end{proof}

Going in the opposite direction, we can take quotients of graded algebras and modules to obtain corresponding filtered superalgebras and supermodules, provided
that we have injective degree 2 shift maps which respect the product and module structures.

\begin{lemma}\label{lemma:shift}
	Let $\A$ be a $\Z$-graded algebra, and $\M$ a $\Z$-graded $\A$-module. Suppose
	futher that we have injective even shift maps
	$\sigma_{A} : \A_{p} \to \A_{p+2}$ and $\sigma_{\M} : \M_{p} \to \M_{p+2}$
	which satisfy the identities
	\begin{align}
	\label{eq:shift-A}
		\sigma_{\A} (ab) &= (\sigma_{A} a)b = a (\sigma_{A} b), \\
	\label{eq:shift-M}
		\sigma_{\M}(am) &= (\sigma_{\A}a)m = a(\sigma_{\M}m)
	\end{align}
	with respect to the product of $a,b\in \A$ and the action of $a\in \A$ on $m \in \M$.
	\begin{enumerate}
	\item
	The quotient $\Asigma$ is a filtered superalgebra.
	\item
	The quotient $\Msigma$ is a filtered supermodule over both $\A$ and $\Asigma$.
	\item
	The formal deformation $\Deft \Asigma$ is a graded algebra isomorphic to $\A$.
	\item
	The formal deformation $\Deft \Msigma$ is a graded $\Deft \Asigma$-module isomorphic to $\M$.
	\end{enumerate}
\end{lemma}

\begin{proof}
	For (1), it follows from (\ref{eq:shift-A}) that $(\sigma - 1)\A$ is a two-sided ideal, as we have
\begin{align*}
	a ( \sigma b - b) &= a (\sigma b) - ab = \sigma(ab) - ab, \\
	( \sigma a - a) b&= (\sigma a)b - ab = \sigma (ab) - ab,
\end{align*}
for $a,b\in A$. Since the product on $\A$ respects the $\Z$-grading, it also respects the filtration given by (\ref{eq:natural-filtration}).
It follows that the product on the quotient $\Asigma$ likewise respects the projected filtration (\ref{eq:filtration-Vprime}) and residual $\Z_{2}$-grading. Therefore $\Asigma$ is a filtered superalgebra.

Similarly, for (2) it follows from (\ref{eq:shift-M}) that $(\sigma-1)\M$ is an $\A$-invariant subspace of $M$, as
$$a (\sigma m - m) = a(\sigma m) - am = \sigma(am) - am$$
for $a\in A$ and $m \in M$.  In addition, since the action of $\A$ on $\M$ respects the gradings, it respects the filtrations (\ref{eq:natural-filtration}), and so the action of $\A$ on $\Msigma$ likewise respects the projected filtration (\ref{eq:filtration-Vprime}) on $\Msigma$ and the residual $\Z_{2}$-grading. Thus $\Msigma$ is a filtered $A$-superalgebra.
Furthermore, the ideal $(\sigma - 1)\A$ acts trivially on $\Msigma$, since by (\ref{eq:shift-M}) we have
$$(\sigma_{A}a - a) m = (\sigma_{\A}a)m - am = \sigma_{\M}(am) - am$$
for $a\in \A$ and $m \in \M$. Therefore, the $\A$-action on $\Msigma$ descends to an $\Asigma$-action on $\Msigma$, making $\Msigma$ a filtered $\Asigma$-supermodule.

For (3) and (4), we note that the map $f$ given by (\ref{eq:f}) is a homomorphism of algebras when applied to $\A$ and a homomorphism of $\A$-modules when applied to $\M$.
\end{proof}

Finally, our main result for this section is the equivalence of filtered supermodules and graded supermodules with injective degree 2 shifts. This is simply a restatement of the above lemmas in the form which we will use in later sections.

\begin{corollary}\label{cor:main}
	Let $A$ be a filtered superalgebra and $\A$ a graded algebra with injective degree 2 shift $\sigma$ satisfying (\ref{eq:shift-A}) which is isomorphic to the formal deformation $\Deft A$. Then the formal deformation construction
	$M \mapsto \Deft M$ and the quotient construction $\M \mapsto \Msigma$ give inverse one-to-one correspondences
	between isomorphism classes of filtered $A$-supermodules and isomorphism
	classes of graded $\A$-modules with injective degree 2 shifts satisfying (\ref{eq:shift-M}).
\end{corollary}

\begin{example}
	Let $\g = \g_{0} \oplus \g_{1}$ be a Lie superalgebra.  We can define an integral filtration on $\g$ by taking $F_{p}\g = \g_{p\bmod 2}$ for $p > 0$ and $F_{p}\g = 0$ for $p \le 0$. This filtration is clearly multiplicative. The corresponding formal deformation $\Deft \g$ is isomorphic to $\g_{1}$ in odd degrees starting with 1 and $\g_{0}$ in even degrees starting with 2. If $\g$ is semi-simple, i.e., $[\g,\g] = \g$, then $\Deft\g$ is the $\Z$-graded Lie superalgebra generated by $\g_{1}t$ and $\g_{0}t^{2}$. We note that the graded super Poincar\'e algebra $\poZ^{1|N}$ is a graded Lie sub-superalgebra of the formal deformation
$\Deft \po^{1|N}$ of the super Poincar\'e algebra.
\end{example}

\begin{remark}
	Mapping our notation to Gerstenhaber's \cite{Ger66}, we have
	$$F_{p}A \longrightarrow F_{-p}A, \qquad
	 s \longrightarrow t^{-2}, \qquad
	 \Deft A \longrightarrow \App A$$
	Also, our filtrations, which generally satisfy $F_{p} = 0$ for $p < 0$, correspond
	to non-positive filtrations in Gerstenhaber's notation. Finally, Gerstenhaber
	defines the filtration on the quotient $\Vsigma$ by taking the image of the the homogeneous graded components $\V_{p}$ under the projection $\pi : \V \to \Vsigma$. In contrast
	we use the $\Z$-grading on $\V$ to define a filtration first on $\V$, and then take the images of the filtration levels $F_{p}\V$ via $\pi$ to obtain the filtration on $\Vsigma$. These two definitions are equivalent by (\ref{eq:two-filtrations}), provided that the shift is injective.
\end{remark}

\section{Filtered Clifford supermodules}

In this section, we apply the general results of the previous section to the correspondence between graded off-shell and filtered on-shell representations.

\begin{definition}\label{def:clifford}
	The \textit{Clifford algebra} $\Cl(N)$ is the real superalgebra generated by the odd Clifford generators
	$\gamma_{1}, \ldots, \gamma_{N}$, subject to the anti-commutation relations
	$$ \{ \gamma_{i},\gamma_{j} \} = 2 \, \delta_{ij}\Id.$$
	The Clifford algebra is a filtered superalgebra with respect to the filtration
	$$F_{p}\bigl( \Cl(N) \bigr) := \Span \bigl\{ \gamma_{i_{1}} \cdots \gamma_{i_{j}} : j \leq p \text{ and } j \equiv p \bmod 2 \bigr\},$$
	given by assigning each of the Clifford generators $\gamma_{1},\ldots,\gamma_{N}$ filtration degree 1.
\end{definition}

In other words, the $p$-th filtration degree of the Clifford algebra consists of everything obtained by taking products of up to $p$ Clifford generators, while also restricting to the $\Z_{2}$-homogeneous subspace of the same parity as $p$. We have canonical isomorphisms:
\begin{align*}
	F_{0}\bigl( \Cl(N) \bigr) &= \R\Id \cong \R, \\
	F_{1}\bigl( \Cl(N) \bigr) &= \Span\{\gamma_{1},\ldots,\gamma_{N}\} \cong \R^{N}, \\
	F_{2}\bigl( \Cl(N) \bigl) &= \R\Id \oplus \Span \bigl\{ \gamma_{i}\gamma_{j} : 1 \leq i < j \leq N \bigr\} \cong \R \oplus \Lambda^{2}(\R^{N}).
\end{align*}
More generally, given an inner product space $\bbS$, we can define the Clifford algebra as the quotient of the free tensor algebra $T^{*}(\bbS)$ by an inhomogeneous even ideal,
$$\Cl(\bbS) := T^{*}(\bbS) \,/\, \bigl( u \otimes v + v \otimes u = 2\,\langle u,v\rangle\,1 \bigr).$$
The $\Z$-grading on the free tensor algebra descends to a $\Z_{2}$-grading on $\Cl(N)$, making it an associative superalgebra. In addition, the natural induced filtration the free tensor algebra given by (\ref{eq:natural-filtration})
descends to a filtration on the quotient $\Cl(N)$, making it a filtered superalgebra. The canonical isomorphisms above then become
$$F_{0}\Cl(\bbS) \cong \R, \qquad
  F_{1}\Cl(\bbS) \cong \bbS, \qquad
  F_{2}\Cl(\bbS) \cong \R \oplus \Lambda^{2}(\bbS).$$
See \cite{ABS64} or \cite{LM89} for further discussion of Clifford algebras.

\begin{lemma}\label{lemma:main}
	The filtered Lie superalgebra homomorphism $\poZ^{1|N} \to \Cl(N)$ given by
	\begin{equation}\label{eq:poincare-clifford}
	Q_{i} \mapsto \gamma_{i}, \qquad H \mapsto 1
	\end{equation}
	induces an isomorphism 
	\begin{equation}\label{eq:poincare-clifford-iso}
	U\bigl(\poZ^{1|N}\bigr)_{H-1} = U\bigl(\poZ^{1|N}\bigr) \,/\, (H-1)U\bigl(\poZ^{1|N}\bigr) \cong \Cl(N)
	\end{equation}
	of filtered superalgebras.
	Conversely, the map
	\begin{equation}\label{eq:poincare-clifford-2}
	Q_{i} \mapsto \gamma_{i}s, \qquad H \mapsto s^{2}
	\end{equation}
	induces a graded algebra isomorphism $U(\poZ^{1|N}) \cong \Deft \Cl(N)$.
\end{lemma}
\begin{proof}
We see that the map given by (\ref{eq:poincare-clifford}) is indeed a Lie superalgebra homomorphism $\po^{1|N} \to \Cl(N)$ by comparing the brackets in Definition~\ref{def:poincare} for the super Poincar\'e algebra and Definition~\ref{def:clifford} for the Clifford algebra. This homomorphism therefore extends to an associative algebra homomorphism $U(\po^{1|N})\to\Cl(N)$ by the universal property of $U(\po^{1|N})$. It is clearly surjective, and its kernel is the ideal
generated by $H-1$, so it induces the isomorphism (\ref{eq:poincare-clifford-iso}). In addition, the map (\ref{eq:poincare-clifford}) preserves the $\Z_{2}$-grading and filtration, so we obtain an isomorphism of filtered superalgebras.

Applying Lemma~\ref{lemma:shift}, it follows from (\ref{eq:poincare-clifford-iso}) that the formal deformation $\Deft\Cl(N)$ is isomorphic to the universal enveloping superalgebra $U(\poZ^{1|N})$ as graded algebras. The map (\ref{eq:poincare-clifford-2}) is a graded homomorphism which takes generators to generators, and so it must therefore be an isomorphism. We note that the composition of the map (\ref{eq:poincare-clifford-2}) with the evaluation homomorphism $\ev_{1} : \Deft\Cl(N) \to \Cl(N)$ given by (\ref{eq:ev1}), is the map (\ref{eq:poincare-clifford}).
\end{proof}

We can now state our main theorem, which in light of the above Lemma is a special case of Corollary~\ref{cor:main}.

\begin{theorem}\label{theorem:main}
	The formal deformation construction $V \mapsto \Deft V$ and the quotient construction $\V \mapsto \V_{H-1}$ give inverse one-to-one correspondences between the isomorphism classes of filtered $\Cl(N)$-supermodules and the isomorphism classes of graded formal off-shell representations of the graded super Poincar\'e algebra $\po^{1|N}_{\Z}$.
\end{theorem}
\begin{proof}
	We established in Lemma~\ref{lemma:main} that the universal enveloping algebra $U(\poZ^{1|N})$ is isomorphic to the formal deformation $\Deft \Cl(N)$ of the Clifford algebra, with the generator $H$ mapping to the shift $s^{2}$. We recall that a formal off-shell
representation of $\poZ^{1|N}$, the shift operator $H$ acts injectively,
and it automatically satisfies the identity (\ref{eq:shift-M}). We can therefore apply Corollary~\ref{cor:main}.
\end{proof}

We now construct explicitly this identification of graded off-shell representations and filtered on-shell representations.  Let $\V$ be a formal graded off-shell representation of the graded super Poincar\'e algebra $\poZ^{1|N}$. Since off-shell representations are finitely generated free $\R[H]$-supermodules, we can write $\V \cong \R[H] \otimes V$, where $\deg_{H} = 2$ and $V = \V/H\V$ is a finite-dimensional $\Z$-graded vector space. Since $V$ is finite-dimensional, it has a minimum degree, which we typically normalize to be either $0$ or $1$, and in particular we require that $V_{p} = 0$ for $p<0$. Identifying $V$ with the subspace $1 \otimes V \subset \R[H] \otimes V \cong \V$, the homogeneous degree subspaces of $\V$ are then of the form:
\begin{align*}
\V_{0} &= V_{0}, & \V_{1}&= V_{1}, \\
\V_{2} &= HV_{0} \oplus  V_{2}, & \V_{3} &= HV_{1} \oplus V_{3}, \\
\V_{2p} &= H^{p}V_{0} \oplus H^{p-1}V_{2} \oplus \cdots \oplus V_{2p}, &
\V_{2p+1} &= H^{p} V_{1}\oplus H^{p-1} V_{3} \oplus \cdots \oplus  V_{2p+1}.
\end{align*}
In general, we have $\V_{p} \cong H\V_{p-2} \oplus V_{p}$.  However, the finite-dimensional graded vector space $V$ must also have a maximal degree $m$, giving us $V_{p} = 0$ for $p > m$. It follows that $\V_{p} = H\V_{p-2}$ for $p > m$. In other words, the grading stabilizes above degree $m$, giving us a stable super vector space with even and odd components:
\begin{align*}
\V_{m-1} &\cong \V_{m+1} \cong \V_{m+3} \cong \cdots \cong \lim_{\longrightarrow}\V_{m-1+2p}, \\
\V_{m} &\cong \V_{m+2} \cong \V_{m+4} \cong \cdots \cong \lim_{\longrightarrow}\V_{m+2p},
\end{align*}
where the isomorphisms are given by the action of $H$.
We note that the super vector space $\V_{m-1}\oplus \V_{m}$ is isomorphic
to the $\Z_{2}$-reduction of the graded vector space $V$. 
The action of the super Poincar\'e algebra on $\V$ descends to an action of the Clifford algebra on the stable space $\V_{m-1}\oplus \V_{m}$, with the action of $H$ descending to the identity operator.  More precisely, the Clifford
generators $\gamma_{1},\ldots,\gamma_{N}\in\Cl(N)$ act by:
$$\gamma_{i} : \V_{m-1} \oplus \V_{m} \xrightarrow{\; Q_{i}\;} \V_{m} \oplus \V_{m+1} \xrightarrow{\left(\begin{smallmatrix}0 & H^{-1} \\ \Id & 0\end{smallmatrix}\right)} \V_{m-1}\oplus\V_{m}.$$
In fact, the stable space $\V_{m-1}\oplus\V_{m}$ is a filtered Clifford supermodule, with filtration
\begin{equation}\label{eq:on-shell-from-off-shell}
F_{p} ( \V_{m-1}\oplus \V_{m} ) := H^{\left\lfloor\frac{m-p}{2}\right\rfloor} \V_{p} \cong \V_{p}.
\end{equation}
In other words, we use the appropriate power of the injection $H$ to identify $\V_{p}$ with a subspace of $\V_{m}\oplus \V_{m+1}$.

Going in the other direction, suppose that $V$ is a finite-dimensional filtered $\Cl(N)$-super\-module. The corresponding formal graded off-shell representation of $\poZ^{1|N}$ is then
\begin{equation}\label{eq:off-shell-from-on-shell}
\V = \bigoplus_{p\in\Z}\V_{p} \text{ with } \V_{p} := F_{p}V.
\end{equation}
The even generator $H$ and supersymmetry generators $Q_{1},\ldots,Q_{N}$ of $\poZ^{1|N}$
then act on $\V$ according to the following commutative diagrams:
$$\xymatrix{
\V_{p} \ar[r]^-{H} \ar@{=}[d] & \V_{p+2} \ar@{=}[d] \\
  F_{p}V \ar@{^(->}[r] & F_{p+2}V
} \qquad
\xymatrix{
\V_{p} \ar[r]^-{Q_{i}} \ar@{=}[d] & \V_{p+1} \ar@{=}[d] \\
  F_{p}V \ar[r]^-{\gamma_{i}} & F_{p+1}V
}
$$
with actions induced by the inclusion of adjacent filtration levels and the Clifford actions of the Clifford generators $\gamma_{1},\ldots,\gamma_{N}\in\Cl(N)$.

Since we identify the action of $H$ with the identity map on the filtered Clifford supermodule $\V_{m-1}\oplus\V_{m}$, we see that it is isomorphic to the formal on-shell representation $\V_{H-1}$. In addition, the formal graded off-shell representation $\V$ given by (\ref{eq:off-shell-from-on-shell}) is clearly isomorphic to $\Deft V$, and all that is missing is the formal parameter $s$ to keep track of the gradings. It then follows from Theorem~\ref{theorem:main} that these two constructions are inverses of each other, which we can see at the level of vector spaces by comparing (\ref{eq:on-shell-from-off-shell}) with (\ref{eq:off-shell-from-on-shell}). We therefore have a one-to-one correspondence between formal graded off-shell representations and formal filtered on-shell representations of one-dimensional supersymmetry.

\section{Bifiltrations}

In this section we consider results analogous to those of Section~\ref{sec:filtrations} and Section~\ref{sec:deformations} where we consider
filtrations and gradings with respect to $\Z \times \Z$.

\begin{definition}
An (increasing) bifiltration on a $\Z_{2}$-bigraded vector space
$$V = V_{00} \oplus V_{01} \oplus V_{10} \oplus V_{01}$$
is a collection of subspaces $F_{p,q}V \subset V_{p\bmod 2,\,q\bmod 2}$
for $p,q\in \Z$ such that
$$F_{p,q}V \subset F_{p+2,q}, \qquad
  F_{p,q}V \subset F_{p,q+2}.$$
We normalize bifiltrations so that $F_{p,q}V = 0$ if $p < 0$ or $q<0$.
\end{definition}

In other words, a bifiltration on a bi-super space is a collection of four separate $\Z\times\Z$ filtrations corresponding to each of the four homogeneous $\Z_{2}\times\Z_{2}$-degree components.
Our positivity condition means that we are considering first quadrant bifiltrations.

\begin{definition}
A bifiltered $\Z_{2}$-bigraded algebra is an algebra $A$ with $\Z_{2}$-bigraded bifiltration satisfying the multiplicative property
	$$F_{p,q}A \cdot F_{m,n}A \subset F_{p+m,q+n}A$$
for $p,q,m,n\in \Z$. A bifiltered $\Z_{2}$-bigraded module over a bifiltered $\Z_{2}$-bigraded algebra $A$ is an $A$-module $M$ with a $\Z_{2}$-bigraded bifiltration satisfying
	$$F_{p,q}A \cdot F_{m,n}M \subset F_{p+m,q+n}M$$
for $p,q,m,n\in \Z$.
\end{definition}

When presented with bifiltrations, we can construct formal deformations involving two parameters rather than one.

\begin{definition}
	Given a (increasingly) bifiltered $\Z_{2}$-bigraded vector space $V$, its formal deformation $\Defst V$ is the subspace of $V[s,t]$ consisting of finite polynomials in two variables of the form
\begin{equation}\label{eq:Defst}
\Defst V := \left\{ \sum_{p,q\in\Z} v_{p,q}\, s^{p}t^{q} \in V[s,t] : v_{p,q}\in F_{p,q}V \right\}.
\end{equation}
The formal deformation $\Defst V$ is naturally $\Z$-bigraded by powers of the variables $s$ and $t$, with homogeneous $\Z$-bidegree components
\begin{equation}\label{eq:bidegree-is-bifiltration}
(\Defst V)_{p,q} = F_{p,q}V\,s^{p}t^{q}
\end{equation}
for $p,q\in\Z$
\end{definition}

Alternatively, we can construct the formal deformation $\Defst V$  in two stages. Given a bifiltered super vector space $V$, we can view it as a single filtered space with respect to the left filtrations degree by taking
the direct limits with respect to the right filtration degree:
$$F_{p}V := \lim_{q\to\infty} F_{p,q}V.$$
The grading on the formal deformation $\Deft V$ corresponds to the left filtration degree on $V$, but $\Deft V$ also admits a residual filtration induced by the right filtration degree. We can therefore take a second formal deformation, and we obtain
$\Defst V = \Def_{t}(\Def_{s}V).$

Going in the reverse direction,
suppose that $\V$ is a $\Z$-bigraded algebra together with two injective even shifts
\begin{equation}\label{eq:sigma-tau-shifts}
\sigma : \V_{p,q} \to \V_{p+2,q}, \qquad
  \tau : \V_{p,q} \to \V_{p,q+2}
\end{equation}
such that
$$\sigma \circ \tau = \tau \circ \sigma : \V_{p,q} \to \V_{p+2,q+2}.$$
Consider the subspace
\begin{align*}
(\sigma -1,\tau-1)\V &:= ( \sigma - 1 ) \V + (\tau -1) \V \\
&= \{ \sigma x - x + \tau y - y : x,y\in \V \}.
\end{align*}
In particular, we note that
\begin{align*}
\sigma \tau z- z &= \sigma (\tau z) - (\tau z)  + \tau z - z \;\,\in (\sigma-1,\tau-1)\V,\\
\sigma z - \tau z &= \sigma z - z + \tau(-z) + (-z) \in  (\sigma-1,\tau-1)\V,
\end{align*}
for any $z\in\V$. We can then define the quotient
\begin{equation}\label{eq:V-sigma-tau}
\V_{\sigma-1,\tau-1} := \V \,/\, (\sigma-1,\tau-1)\V
\end{equation}
Performing this quotient in two steps, we have equivalently
$$\V_{\sigma-1,\tau-1} = (\V_{\sigma-1})_{\tau-1}.$$
where we note that $(\sigma - 1)\V$ is invariant under $\tau$, since
$$\tau ( \sigma x - x ) = \tau\sigma x - \tau x = \sigma (\tau x) - (\tau x).$$
In addition, the natural bifiltration on $\V$ given by
$$F_{p,q}\V := \bigoplus_{\substack{j \leq p\\j\equiv p\bmod 2}}\,\bigoplus_{\substack{k \leq q\\k\equiv q\bmod 2}}\V_{j,k},$$
descends to a bifiltration on $\V_{\sigma-1,\tau-1}$ given by
$$F_{p,q}\V_{\sigma-1,\tau-1} := F_{p,q}\V \,/\, \bigl( (\sigma-1)F_{p-2,q}\V + (\tau - 1)F_{p,q-2}\mathcal{V}\bigr).$$
By the analogue of the telescoping construction described in (\ref{eq:two-filtrations}), we obtain
\begin{equation}\label{eq:bifiltration-is-bidegree}
F_{p,q}\V_{\sigma-1,\tau-1} \cong \V_{p,q}.
\end{equation}
Comparing (\ref{eq:bidegree-is-bifiltration}) with (\ref{eq:bifiltration-is-bidegree}), we see that the formal deformation construction (\ref{eq:Defst}) and the quotient construction (\ref{eq:V-sigma-tau}) are inverses of one another. We have therefore established the following lemma:

\begin{lemma}
If $V$ is a bifiltered super vector space, then the bifiltered super vector space $(\Defst V)_{s^{2}-1,t^{2}-1}$ is canonically isomorphic to $V$.  Conversely, if $\V$ is a bigraded vector space with a commuting pair of injective degree 2 shifts $\sigma,\tau$ as in (\ref{eq:sigma-tau-shifts}), then $\Defst (\V_{\sigma-1,\tau-1})$ is isomorphic to $\V$, identifying the shifts $(\sigma,\tau)$ with $(s^{2},t^{2})$.
\end{lemma}

We now reprise Lemma~\ref{lemma:def1} and Lemma~\ref{lemma:shift} 
in the bifiltered and bigraded setting. We state these results without proof, as our arguments from Section~\ref{sec:deformations} generalize
with only minor modifications. We leave the details to the reader.

\begin{lemma}\label{lemma:bi-def1}
	Let $A$ be a bifiltered superalgebra, and let $M$ be a bifiltered $A$-supermodule.
	\begin{enumerate}
		\item
		The formal deformation $\Defst A$ is a $\Z$-bigraded subalgebra of $A[s,t]$.
		\item
		The formal deformation $\Deft M \subset M[s,t]$ is a $\Z$-bigraded module
		over $\Defst A$.
		\item
		The quotient $(\Defst A)_{s^{2}-1,t^{2}-1}$ is a bifiltered superalgebra
		canonically isomorphic to $A$.
		\item
		The quotient $(\Defst M)_{s^{2}-1,t^{2}-1}$ is a bifiltered $(\Defst A)_{s^{2}-1,t^{2}-1}$-supermodule
		canonically isomorphic to $M$.
	\end{enumerate}
\end{lemma}

\begin{lemma}\label{lemma:bi-shift}
	Let $\A$ be a $\Z$-bigraded algebra, and $\M$ a $\Z$-bigraded $\A$-module. Suppose
	futher that we have commuting pairs of injective even shift maps
	$\sigma_{A} : \A_{p,q} \to \A_{p+2,q}$, $\tau_{\A} : \A_{p,q}\to\A_{p,q+2}$ and $\sigma_{\M} : \M_{p,q} \to \M_{p+2,q}$, $\tau_{\M} : \M_{p,q} \to \M_{p,q+2}$
	which satisfy the identities
	\begin{align}
	\label{eq:bi-shift-A}
		\sigma_{\A} (ab) &= (\sigma_{A} a)b = a (\sigma_{A} b), &
		\tau_{\A} (ab) &= (\tau_{A} a)b = a (\tau_{A} b),
		\\
	\label{eq:bi-shift-M}
		\sigma_{\M}(am) &= (\sigma_{\A}a)m = a(\sigma_{\M}m) &
		\tau_{\M}(am) &= (\tau_{\A}a)m = a(\tau_{\M}m)
	\end{align}
	with respect to the product of $a,b\in \A$ and the action of $a\in \A$ on $m \in \M$.
	\begin{enumerate}
	\item
	The quotient $\A_{\sigma-1,\tau-1}$ is a bifiltered superalgebra.
	\item
	The quotient $\M_{\sigma-1,\tau-1}$ is a bifiltered supermodule over both $\A$ and $\A_{\sigma-1,\tau-1}$.
	\item
	The formal deformation $\Defst \A_{\sigma-1,\tau-1}$ is a bigraded algebra isomorphic to $\A$.
	\item
	The formal deformation $\Defst \M_{\sigma-1,\tau-1}$ is a bigraded $\Defst \A_{\sigma-1,\tau-1}$-module isomorphic to $\M$.
	\end{enumerate}
\end{lemma}

Finally, we conclude this section with the bifiltered and bigraded version of Corollary~\ref{cor:main}, summarizing the results which we will need in our application in the next section.

\begin{corollary}\label{cor:bi-main}
	Let $A$ be a bifiltered superalgebra and $\A$ a bigraded algebra with a commuting pair of injective degree 2 shifts $\sigma,\tau$ satisfying (\ref{eq:bi-shift-A}) which is isomorphic to the formal deformation $\Defst A$. Then the formal deformation construction
	$M \mapsto \Defst M$ and the quotient construction $\M \mapsto \M_{\sigma-1,\tau-1}$ give inverse one-to-one correspondences
	between the isomorphism classes of bifiltered $A$-supermodules and the isomorphism
	classes of bigraded $\A$-modules with commuting pairs of injective degree 2 shifts satisfying (\ref{eq:bi-shift-M}).
\end{corollary}

\section{Two-dimensional supersymmetry}

In $1+1$ dimensions, the vector representation of $\Spin(1,1)$ decomposes as the direct sum of two complementary light-cone representations $\R^{1,1} \cong \R^{+} \oplus \R^{-}$, known as the left-movers and right-movers. Likewise, there are two chiral spin representations $\bbS^{+}$ and $\bbS^{-}$, with $\Sym^{2}(\bbS^{\pm}) = \R^{\pm}$. When constructing the 1+1-dimensional super Poincar\'e algebra, we can decompose the supersymmetry generators into chiral generators squaring to multiples of $H+P$ and anti-chiral generators squaring to multiples of $H-P$.

\begin{definition}
	The $d=2$, $(p,q)$-extended super Poincar\'e algebra is the real Lie superalgebra
	$$\po^{1,1|p,q} = \R L \oplus \Span\{H,P\} \oplus \Pi\Span\{Q^{+}_{1},\ldots,Q^{+}_{p},Q^{-},\ldots,Q^{-}_{q}\}$$
with non-vanishing super brackets
$$[L,H\pm P] = \pm 2\,(H\pm P), \qquad
  [L,Q^{\pm}_{i}] = \pm Q^{\pm}_{i}, \qquad
  \{Q^{\pm}_{i},Q^{\pm}_{j}\} = 2 \,\delta_{ij}\,(H\pm P).$$
It is a $\Z$-graded Lie superalgebra, with degrees given by
\begin{align*}
\deg L = 0, \qquad \deg H = \deg P = 2, \qquad
\deg Q_{i}^{\pm} = 1.
\end{align*}
(We will use the same notation for the $\Z$-graded and $\Z_{2}$-graded version.)
\end{definition}

We obtain an alternative description of the $d=2$ super Poincar\'e algebra by explicitly splitting up the left-movers and right movers as follows:
\begin{equation}\label{eq:d2-decomposition}
\po^{1,1|p,q} \cong \R L \mathbin{\tilde{\oplus}} \bigl( \po^{1|p} \oplus \po^{1|q} \bigr).
\end{equation}
Here, the chiral and anti-chiral summands are
\begin{align}
\label{eq:p-plus}
\po^{1|p} &\cong \R(H+P) \oplus \Pi\Span\{Q_{1}^{+},\ldots,Q^{+}_{p}\},\\
\label{eq:p-minus}
\po^{1|q} &\cong \R(H-P) \oplus \Pi\Span\{Q_{1}^{-},\ldots,Q^{-}_{q^{-}}\},
\end{align}
respectively.
The degree $0$ generator $L$ acts as an auxiliary grading operator, with eigenvalue called the helicity. Given a $\Z$-graded representation of $\po^{1,1|p,q}$, such as the adjoint action on itself, we can combine the degree and helicity into a $\Z$-bigrading given by
\begin{equation}\label{eq:deg-pm}
\deg_{+}x = \frac{ \deg x + h }{2}, \qquad
  \deg_{-}x = \frac{ \deg x - h}{2}.
\end{equation}
for a homogeneous degree $L$-eigenvector $x$ with $L(x) = h(x)$.
We note that $\deg_{+}$ is the chiral degree with respect to the $\po^{1|p}$ summand (\ref{eq:p-plus}), while $\deg_{-}$ is the anti-chiral degree with respect to the $\po^{1|q}$ summand (\ref{eq:p-minus}). Conversely, for a homogeneous bigraded element $x$ of bidegree $(\deg_{+}x,\deg_{-}x) = (m,n)$, we have
$$\deg x = m+n, \qquad L(x) = (m-n)x,$$
and in particular the generator $L$ has bidegree $(0,0)$.

\begin{definition}
A \textit{formal off-shell representation} of the $d=2$ super Poincar\'e algebra is a $\Z$-graded representation of $\po^{1,1|p,q}$ on a finitely generated, free $\R[H,P]$-supermodule. Furthermore, we
require that it obeys spin-statistics, with helicities $h \equiv \deg\pmod 2$.
\end{definition}

In light of our discussion above of the bigrading on the super Poincar\'e algebra, we obtain the following lemma:

\begin{lemma}\label{lemma:po-plus-po}
Via the decomposition (\ref{eq:d2-decomposition}), formal off-shell representations of $\po^{1,1|p,q}$ are equivalent to $\Z$-bigraded representations of the direct sum $\po^{1|p} \oplus \po^{1|q}$ on finitely generated, free $\R[H+P] \otimes \R[H-P]$-supermodules.
\end{lemma}

\begin{proof}
All we need to verify is that the bigrading $(\deg_{+},\deg_{-})$ is indeed integral. However, this follows from the spin-statistics condition and (\ref{eq:deg-pm}).
\end{proof}

In other words, we consider $\Z$-bigraded representations $\V$ with two injective even shifts
$$H+P : \V_{m,n} \to \V_{m+2,n}, \qquad H-P : \V_{m,n}\to \V_{m,n+2}.$$
Using the techniques of the previous sections, we obtain the following correspondence:

\begin{definition}
The twisted tensor product $A \mathbin{\tilde{\otimes}} B$ of two superalgebras $A$ and $B$ is the superalgebra whose underlying vector space is the conventional tensor product, which has $\Z_{2}$-bidegree given by
$$\deg (a \mathbin{\tilde{\otimes}} b) = (\deg a,\deg b)$$
for $a\in A$ and $b\in B$, and which has multiplication
$$( a_{1} \mathrel{\tilde{\otimes}} b_{1} )
 ( a_{2} \mathrel{\tilde{\otimes}} b_{2} )
 := (-1)^{|b_{1}|\,|a_{2}|} a_{1} a_{2} \mathrel{\tilde{\otimes}} b_{1}b_{2}$$
for homogeneous elements $a_{1},a_{2}\in A$ and $b_{1},b_{2}\in B$. If in addition $A$ and $B$ are filtered superalgebras, then $A \mathbin{\tilde{\otimes}} B$ is a bifiltered superalgebra, with
\begin{equation}\label{eq:bifiltered-product}
F_{m,n} ( A \mathbin{\tilde{\otimes}} B )
= F_{m} A \mathbin{\tilde{\otimes}} F_{n} B
\end{equation}
for $m,n\in \Z$.
\end{definition}

Our main use for such twisted products is the following definition:

\begin{definition}
A bifiltered Clifford algebra is the twisted tensor product
$$\Cl(p+q) := \Cl(p) \mathbin{\tilde{\otimes}} \Cl(q)$$
of two Clifford algebras, with bifiltration given by (\ref{eq:bifiltered-product}).
\end{definition}

Alternatively, we could define $\Cl(p+q)$ to be the Clifford algebra determined by two sets of Clifford generators:
\begin{align*}
\gamma_{1}^{+},\ldots,\gamma_{p}^{+} \text{ with filtration degree }(1,0), \\
\gamma_{1}^{-},\ldots,\gamma_{q}^{-} \text{ with filtration degree }(0,1).
\end{align*}
It is well known (see, for example, \cite{LM89}) that $\Cl(p+q) \cong \Cl(p) \mathbin{\tilde{\otimes}} \Cl(q)$ as superalgebras.

\begin{theorem}\label{theorem:bi-main}
	The formal deformation construction $V \mapsto \Defst V$ and the quotient construction $\V \mapsto \V_{H+P-1,\,H-P-1}$ give inverse one-to-one correspondences between the isomorphism classes of formal graded off-shell representations of $\po^{1,1|p,q}$ and the isomorphism classes of bifiltered $\Cl(p+q)$-supermodules.
\end{theorem}

\begin{proof}
By Lemma~\ref{lemma:po-plus-po}, formal graded off-shell representations of $\po^{1,1|p,q}$ are equivalent to bigraded representations of the direct
sum $\po^{1|p}\oplus\po^{1|q}$.  Considering universal enveloping superalgebras, we obtain an isomorphism of graded superalgebras
\begin{equation*}\begin{split}
U \bigl( \po^{1|p}\oplus\po^{1|q} \bigr)
&\cong U\bigl( \po^{1|p} \bigr) \mathbin{\tilde{\otimes}} U\bigl( \po^{1|q} \bigr) \\
&\cong \bigl( \Deft \Cl(p)\bigr) \mathbin{\tilde{\otimes}} \bigl(\Def_{t} \Cl(q)\bigr) \\
&\cong \Defst \left( \Cl(p) \mathbin{\tilde{\otimes}} \Cl(q) \right)
\cong \Defst \Cl(p+q),
\end{split}\end{equation*}
using Lemma~\ref{lemma:main}. Here, the degree $2$ generators $H+P$ and $H-P$ in the universal enveloping superalgebra map to the shifts $s^{2}$ and $t^{2}$ in the formal deformation $\Defst$. The commuting pair of shifts $H+P$ and $H-P$ act injectively on formal off-shell representations, and their actions automatically satisfy the identities (\ref{eq:bi-shift-M}). We obtain our result by applying Corollary~\ref{cor:bi-main}.
\end{proof}

\section{Conclusions}

In light of Theorem~\ref{theorem:main} and Theorem~\ref{theorem:bi-main}, the problem of classifying off-shell representations of one-dimensional and two-dimensional supersymmetry becomes the problem of classifying filtered and bifiltered Clifford supermodules.  Even though Clifford modules and supermodules have been a part of the mathematical literature for over 40 years (see \cite{ABS64}), most of the standard references mention the natural filtration on the Clifford algebra only in passing (see \cite{LM89}).  To our knowledge, there has not yet been any systematic study of filtered Clifford supermodules.

Given a Clifford supermodule $V = V_{0} \oplus V_{1}$, a filtration is a pair of flags:
$$F_{0}V \subset F_{2} V \subset \cdots \subset V_{0}, \qquad
  F_{1}V \subset F_{3} V \subset \cdots \subset V_{1},$$
with the added property that the Clifford generators $\gamma_{1},\ldots,\gamma_{N}$ act according to
\begin{equation}\label{eq:gamma-fpV}
\gamma_{i} \cdot F_{p}V \subset F_{p+1} V.
\end{equation}
What are the invariants of such filtrations? The most obvious invariant is the sequence of dimensions of each of the filtration levels:
$$\left(\dim F_{0}V,\, \dim F_{1}V,\, \dim F_{2} V,\, \dim F_{3}V,\, \ldots \right).$$
Alternatively, one could list the sequence of dimensions of the associated graded algebra $\Gr V = \bigoplus_{p\in\Z}F_{p} / F_{p-2}$, giving us
\begin{equation}\label{eq:toppan-classify}
\left(\dim F_{0}V,\, \dim F_{1}V,\, \dim F_{2}V - \dim F_{0} V,\, \dim F_{3}V - \dim F_{1} V,\ldots \right).
\end{equation}
Indeed, reference \cite{Toppan05} discusses the invariant (\ref{eq:toppan-classify}) in the context of classifying irreducible representations appearing in $N$-extended supersymmetric quantum mechanics.

On the other hand, a simple dimension count of the form (\ref{eq:toppan-classify}) does not completely classify off-shell representations of one-dimensional supersymmetry. For example in the $N=4$ case, one can consider the $\Cl(4)$-supermodule
$V=\Lambda^{*}(\R^{4})$. There are then two natural ways to construct
a filtration.  The first is to take the natural filtration corresponding to exterior algebra degrees, with the flags
\begin{equation}\label{eq:14641}
 \Lambda^{0} \,\subset\, \Lambda^{0} \oplus \Lambda^{2} \,\subset
   \Lambda^{0} \oplus \Lambda^{2} \oplus \Lambda^{4} = V_{0}, \quad
   \Lambda^{1} \,\subset\, \Lambda^{1} \oplus \Lambda^{3} = V_{1}.
\end{equation}
For a second filtration, we can use the Hodge star operator to 
decompose the exterior algebra as $\Lambda^{*} = \Lambda^{*}_{+} \oplus \Lambda^{*}_{-}$ into a sum of self-dual and anti-self-dual forms. We can then consider the filtration given by the flags
\begin{equation}\label{eq:143-341}
\Lambda^{0}_{+} \,\subset\, \Lambda^{0}_{+} \oplus \Lambda^{2}_{+} \oplus \Lambda^{2}_{-} \,\subset\, \Lambda^{0}_{+}\oplus \Lambda^{2}_{+}\oplus \Lambda^{2}_{-} \oplus \Lambda^{0}_{-} = V_{0},
\quad
\Lambda^{1}_{+} \,\subset\, \Lambda^{1}_{+} \oplus \Lambda^{1}_{-} = V_{1}.
\end{equation}
For both of the filtrations (\ref{eq:14641}) and (\ref{eq:143-341}), the dimension-counting invariant given by (\ref{eq:toppan-classify}) is $(1, 4, 6, 4, 1)$. However, the filtration  (\ref{eq:14641}) is indecomposable as a filtered Clifford supermodule, while
the filtration (\ref{eq:143-341}) decomposes as the direct sum of a self-dual component with dimensions $(1, 4, 3, 0, 0)$ and an anti-self-dual component with dimensions $(0, 0, 3, 4, 1)$.

Even if we require the original Clifford supermodule to be irreducible, we still find that the dimension-counting invariant (\ref{eq:toppan-classify}) is insufficient to completely classify filtered Clifford supermodules. Starting with the unique irreducible real $\Z_{2}$-graded spin representation of $\Cl(5)$, we have found a family of distinct filtrations, or equivalently a family of distinct off-shell representations, all with the invariant (\ref{eq:toppan-classify}) given by the dimensions $(2,8,6)$. On the other hand, we can show that for $N \leq 4$,
the invariant (\ref{eq:toppan-classify}) does indeed completely
classify the filtrations on an irreducible $\Cl(4)$-supermodule.
We save these topics for a future paper.

To classify these filtrations, one must also take into account the spaces
\begin{equation}\label{eq:fp-mod-Sfp}
F_{p}V \,/\, \Span\{\gamma_{1},\ldots,\gamma_{N}\} \cdot F_{p-1}V.
\end{equation}
In other words, at each filtration degree, one must include the subspaces given by the Clifford action (\ref{eq:gamma-fpV}), but one is then free to add any complementary subspace of the appropriate parity. Considering the dimensions of the spaces (\ref{eq:fp-mod-Sfp}) is sufficient to distinguish the filtrations (\ref{eq:14641}) and (\ref{eq:143-341}), since the filtration (\ref{eq:14641}) has the sequence of dimensions $(1,0,0,0,0)$ while the filtration (\ref{eq:143-341}) has to the sequence of dimensions $(1,0,3,0,0)$. For the family of $(2,8,6)$ filtrations of the irreducible $N=5$ Clifford supermodules described above, the dimensions of the subspaces (\ref{eq:fp-mod-Sfp}) jump from $(2,0,0)$ to $(2,2,0)$. To fully classify these filtrations, we need to consider not only the dimensions of the subspaces (\ref{eq:fp-mod-Sfp}), but also how they are related via the Clifford action.

In \cite{Gates04}, Faux and Gates introduced a combinatorial notation for off-shell representations of one-dimensional supersymmetry, using Coxeter graphs whose vertices correspond to a suitable basis of the Clifford supermodule, and whose edges correspond to the actions of the Clifford generators. The filtration (or corresponding engineering dimension) is then denoted either by directing the edges, or equivalently by placing the vertices at the corresponding height in the graph. In \cite{DFGHIL1}, we prove that the possible height assignments on such graphs are completely determined by the set of sources (or sinks). However, the choices of sources (or sinks) correspond precisely to the possible subspaces (\ref{eq:fp-mod-Sfp}).

Finally, one can ask how these methods may be applied to the classification of off-shell representations of higher-dimensional supersymmetry. In one and two dimensions, the filtration or bifiltration on a Clifford supermodule gives us precisely the information we need to reconstruct how the Minkowski space translations act on the corresponding off-shell representation. In higher dimensions, we do not obtain tri-filtrations or $d$-filtrations, as they violate Lorentz invariance. Our arguments are possible in two dimensions only because $\R^{1,1}$ decomposes as the sum of two one-dimensional light-cone Lorentz invariant subspaces. In higher dimensions, Minkowski space $\R^{1,d-1}$ is irreducible. On the other hand, it should be possible to use the Lorentz invariance to our advantage, and we conjecture that off-shell representations of supersymmetry in any dimension correspond precisely to filtered (or possibly bifiltered) Clifford supermodules possessing an additional compatible Lorentz action. This is the mathematical version of the claim in \cite{Gates02a} that off-shell supersymmetric theories can be classified in terms of their dimensional reduction ``shadows'' in one (or possibly two) dimension.


\begin{thebibliography}{1}

\bibitem{ABS64}
M.~F. Atiyah, R.~Bott, and A.~Shapiro.
\newblock Clifford modules.
\newblock {\em Topology}, 3(suppl.~1):3--38, 1964.

\bibitem{DFGHIL1}
C.~F. Doran, M.~G. Faux, S.~J. Gates, Jr., T.~Hubsch, K.~M. Iga, and G.~D.
  Landweber.
\newblock On graph-theoretic identifications of {A}dinkras, supersymmetry
  representations and superfields.
\newblock December 2005, math-ph/0512016.

\bibitem{Gates04}
M.~G. Faux and S.~J. Gates, Jr.
\newblock Adinkras: A graphical technology for supersymmetric representation
  theory.
\newblock 2004, hep-th/0408004.

\bibitem{Freed99}
D.~S. Freed.
\newblock {\em Five lectures on supersymmetry}.
\newblock American Mathematical Society, Providence, RI, 1999.

\bibitem{Gates02}
S.~J. Gates, Jr., W.~Linch, J.~Phillips, and L.~Rana.
\newblock The fundamental supersymmetry challenge remains.
\newblock {\em Gravit. Cosmol.}, 8(1-2):96--100, 2002, hep-th/0109109.
\newblock Special issue dedicated to the centennial of Tomsk State Pedagogical
  University, arXiv:.

\bibitem{Gates02a}
S.~J. Gates, Jr., W.~D. Linch, III, and J.~Phillips.
\newblock When superspace is not enough.
\newblock 2002, hep-th/0211034.

\bibitem{Ger66}
M.~Gerstenhaber.
\newblock On the deformation of rings and algebras. {II}.
\newblock {\em Ann. of Math.}, 84:1--19, 1966.

\bibitem{Toppan05}
Z.~Kuznetsova, M.~Rojas, and F.~Toppan.
\newblock Classification of irreps and invariants of the {$N$}-extended
  supersymmetric quantum mechanics.
\newblock November 2005, hep-th/0511274.

\bibitem{LM89}
H.~B. Lawson and M.-L. Michelson.
\newblock {\em Spin Geometry}, volume~38 of {\em Princeton Mathematical
  Series}.
\newblock Princeton University Press, Princeton, N.J., 1989.

\end{thebibliography}

\def\cprime{$'$}

 \end{document}